\documentclass[aps,prb,reprint,superscriptaddress,amsmath,amssymb]{revtex4-2}
\usepackage[utf8]{inputenc} 

\DeclareUnicodeCharacter{0307}{} 
\usepackage[table]{xcolor}
\usepackage{xcolor}
\usepackage{graphicx}
\DeclareGraphicsExtensions{.png,.pdf,.tif,.jpeg}
\DeclareUnicodeCharacter{03BA}{\ensuremath{\kappa}}

\usepackage{parskip}  
\usepackage{siunitx}
\usepackage{amsmath}

\begin{document}

\title{Evidence for Multiband Superconductivity in 2H-NbSeS}

\author{K. Yadav}
\author{M. Lamba}
\affiliation{School of Physical Sciences, Jawaharlal Nehru University, New Delhi-110067, India}

\author{K. Bhattacharya}
\author{M. Majumder}
\affiliation{Department of Physics, Shiv Nadar Institution of Eminence, Gautam Buddha Nagar, UP 201314, India}

\author{S. Patnaik}
\email{spatnaik@jnu.ac.in}
\affiliation{School of Physical Sciences, Jawaharlal Nehru University, New Delhi-110067, India}

\date{\today}

\begin{abstract}

The nature of superconductivity in 2H-NbSe$_2$ has generated sustained debate in the recent past. While angle resolved photoemission spectroscopy data have been interpreted as evidence for multiband superconductivity, the data from scanning tunneling microscope experiments relate to strongly anisotropic single-band superconductivity. In the later case, the charge density wave (CDW) order mimics the multigap character. Because the CDW reconstructs the Fermi surface and modifies the superconducting gap distribution, disentangling intrinsic multiband pairing from CDW-related effects is challenging. To address this issue, we investigate single-crystalline 2H-NbSeS, a mixed-chalcogen analogue of 2H-NbSe$_2$ in which random Se/S substitution suppresses long-range CDW order while preserving the layered crystal structure $P6_3/mmc$. The material becomes superconducting below $T_c \approx 6.0$~K with moderate magnetic anisotropy $\gamma=\xi_{ab}/\xi_c \approx 3.1$. The upper critical field $H_{c2}(T)$ exhibits a pronounced upward curvature that cannot be described within a single-band framework but is well captured by a dirty-limit two-band model with a large diffusivity ratio. This indicates strong band-dependent scattering. The in-plane upper critical field $H_{c2}^{\parallel ab}(0)\approx14.5$~T exceeds the weak-coupling Pauli limit. Measurements of the lower critical field, superfluid density, and electronic specific heat are consistent with an interpretation of a fully gapped superconducting state with two nodeless gaps of different magnitudes.

\end{abstract}


\maketitle

\section{Introduction}

Multiband superconductivity occurs when multiple electronic bands contribute to condensate pairing~\cite{zehetmayer2013review,kruchinin2016multiband}. Under such a scenario, each band can develop a superconducting gap with a different magnitude and anisotropy. In such cases, the mechanism of superconductivity cannot be described within conventional single-band theories~\cite{zehetmayer2013review,kruchinin2016multiband,bouquet2001specific,boaknin2003heat}. In general, such behavior often manifests through enhanced upper critical fields, unconventional thermodynamic properties, and complex vortex structures~\cite{zehetmayer2013review,kruchinin2016multiband,bouquet2001specific,boaknin2003heat}. Because of peculiar band structure, layered transition-metal dichalcogenides (TMDs) provide an ideal platform to investigate such phenomena. Their transition-metal $d$-derived electronic structure naturally produces multiple Fermi-surface sheets~\cite{fletcher2007penetration,inosov2008fermi,wilson1969transition,rossnagel2011origin}. In addition, strong spin-orbit coupling (SOC) and competing charge-density-wave (CDW) order can significantly influence the superconducting state~\cite{wilson1969transition,aase2023constrained,kim2021experimental,luo2020possible}.

Among TMDs, 2H-NbSe$_2$ ($T_c \approx 7.2$~K) is one of the most extensively studied superconductors and an ideal system for investigating multiband superconductivity. Angle-resolved photoemission spectroscopy (ARPES) and thermal conductivity measurements have revealed Fermi-surface-dependent superconducting gaps in single-crystalline 2H-NbSe$_2$~\cite{alshemi2025two,hanaguri2026modulation,yokoya2001fermi,boaknin2003heat}. The largest gaps occur on the Nb-$4d$-derived sheets around the $K$ points, while smaller gaps are observed on other Fermi-surface sheets. The Se-$4p$ pocket centered at $\Gamma$ exhibits a very small superconducting gap~\cite{yokoya2001fermi}. First-principles density functional theory (DFT) calculations have independently confirmed this multi-sheet Fermi surface topology of 2H-NbSe$_2$. These calculations reveal three distinct Fermi-surface sheets; two double-walled Nb-$4d$-derived cylinders centered at the $\Gamma$-$A$ and $K$-$H$ lines of the Brillouin zone, and a small Se-$4p_z$-derived pancake pocket at $\Gamma$~\cite{rossnagel2001fermi,johannes2006fermi}. The Nb-$4d$-derived sheets carry the dominant electron-phonon coupling and are primarily responsible for the superconducting condensate~\cite{johannes2006fermi}. These observations have been interpreted as evidence for multiband superconductivity in 2H-NbSe$_2$. However, an alternative description in terms of a strongly anisotropic nodeless single-band $s$-wave state has also been proposed~\cite{sanna2022real,rahn2012gaps,yasuzuka2020highly, kiss2007charge}. 2H-NbSe$_2$ also develops an incommensurate CDW order below $T_{\mathrm{CDW}} \approx 33$~K~\cite{moncton1977neutron}. The CDW reconstructs portions of the Fermi surface and modifies the momentum dependence of the superconducting gap. These CDW-driven effects can mimic a two-gap structure and are therefore it is difficult to separate from intrinsic multiband  pairing~\cite{sanna2022real,rahn2012gaps,yasuzuka2020highly}. Consequently, the detailed nature of the superconducting gap structure in 2H-NbSe$_2$ remains a topic of ongoing debate, particularly regarding whether it reflects genuine multiband pairing or strong single-band gap anisotropy~\cite{yokoya2001fermi,boaknin2003heat,sanna2022real,yasuzuka2020highly}.

A useful approach to disentangle the effects of CDW order from the intrinsic superconducting gap structure is to study systems in which CDW order is absent while the electronic structure near the Fermi level is expected to remain preserved~\cite{li2017superconducting,Liu2016nature}. Introducing substitutional disorder provides an effective route to achieve this~\cite{li2017superconducting}. Random replacement of one chalcogen by another weakens the long-range periodicity required for CDW formation while largely preserving the layered crystal structure~\cite{li2017superconducting,Liu2016nature}. Mixed-chalcogen compounds therefore provide a valuable platform for studying the intrinsic superconducting gap structure in the absence of CDW order.

In this context, NbSe$_{2-x}$S$x$ is a particularly suitable system. Partial substitution of Se by S introduces intrinsic chalcogen-site disorder that suppresses CDW order while largely preserving the layered crystal structure~\cite{cho2018using,moreno2025gapless,tsuppayakorn2021effect}. This allows the superconducting gap structure to be investigated without the additional complication of CDW-induced Fermi-surface reconstruction. Among the NbSe$_{2-x}$S$x$ series, the equimolar composition 2H-NbSeS ($x = 1$) corresponds to the maximum configurational disorder on the chalcogen sublattice, since the configurational entropy is maximized when Se and S occupy the available sites with equal probability~\cite{kittel1971thermal}. In general, long-range ordered states such as CDWs are particularly sensitive to random substitutional disorder because they rely on coherent periodic modulations of both the electronic density and the lattice~\cite{morosan2006superconductivity,fang2005fabrication}. Consequently, increasing configurational disorder tends to destabilize long-range CDW order by disrupting the coherence required for its formation. As the composition with the highest configurational entropy in the NbSe$_{2-x}$S$_x$ series, 2H-NbSeS is therefore expected to exhibit the strongest disorder-induced suppression of CDW order, providing a favorable platform for investigating the intrinsic superconducting state.

In this work, we report on the synthesis and comprehensive superconducting characterization of single-crystalline 2H-NbSeS. Transport measurements reveal that 2H-NbSeS becomes superconducting below $T_c \approx 6.0$~K with no signature of CDW ordering. The temperature dependence of the upper critical field $H_{c2}(T)$ exhibits a pronounced upward curvature that deviates from a conventional single-band framework. Instead, the data are well captured by a dirty-limit two-band 
model with a large diffusivity imbalance. Furthermore, the in-plane upper critical field $H_{c2}^{\parallel ab}(0)$ exceeds the weak-coupling Pauli limit, pointing to superconducting behavior beyond the single-band description. Additional evidence from measurements of the lower critical field $H_{c1}(T)$, magnetic penetration depth, and electronic specific heat consistently supports a fully gapped superconducting state with two nodeless superconducting gaps of significantly different magnitudes.

\begin{figure}
\centering
\includegraphics[width=\columnwidth]{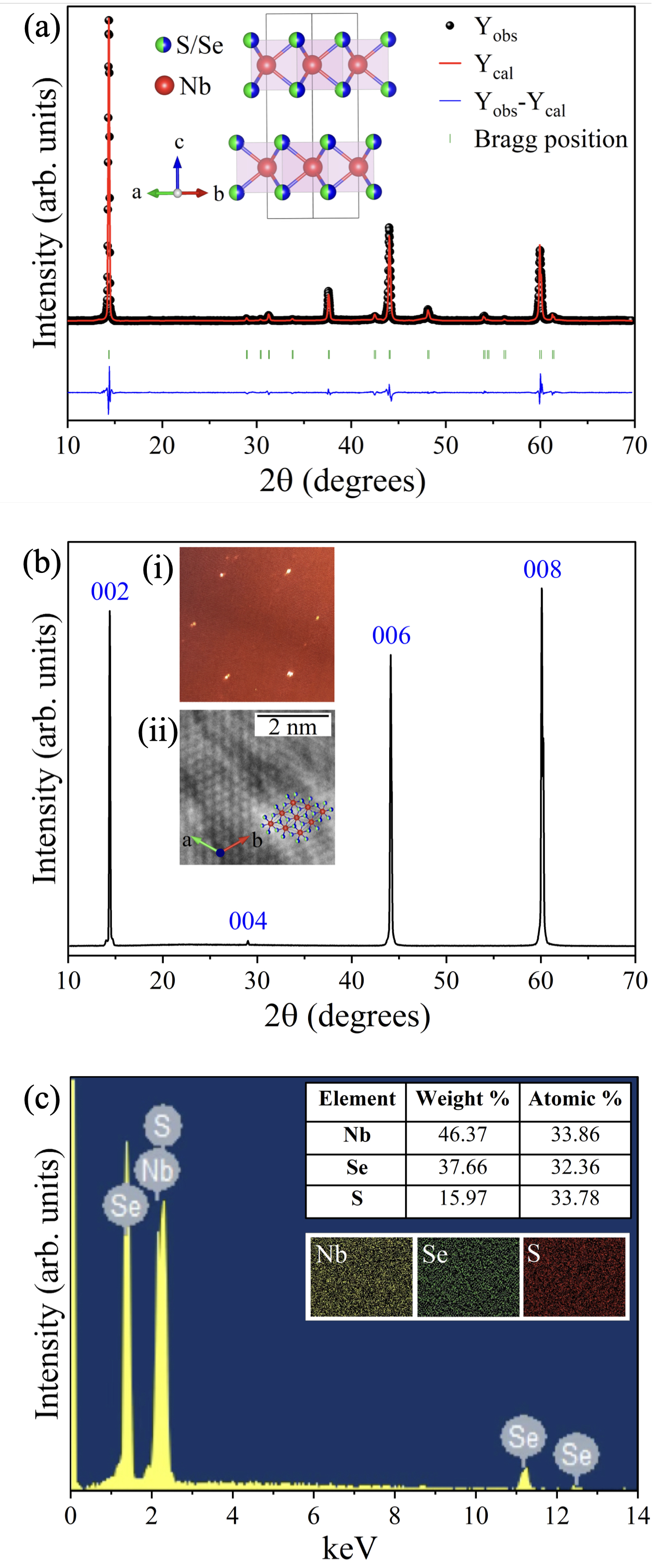}
\caption{(a) Rietveld refinement of the PXRD pattern of 2H-NbSeS; the inset shows the crystal structure of the unit cell. (b) XRD pattern of a NbSeS single crystal showing only $(00l)$ reflections. Inset (i) shows the Laue back-reflection pattern and inset (ii) the HRTEM image revealing the atomic lattice of 2H-NbSeS, overlaid with the simulated crystal structure projected along the $ab$-plane. (c) EDX spectrum recorded from a 2H-NbSeS single crystal, with the table summarizing the mean atomic composition of the constituent elements. The insets present the corresponding elemental distribution maps for the selected region.}
\label{fig:xrd}
\end{figure}

\section{Experimental Techniques}

Single crystals of NbSeS were grown using a modified chemical vapor transport (CVT) method. Stoichiometric amounts of high-purity Nb (99.90\%), Se (99.99\%), and S (99.98\%) powders were thoroughly mixed, pelletized, and sealed in an evacuated quartz ampule ($\sim 10^{-3}$ mbar) with iodine ($\sim 5$ mg/cm$^3$) as the transport agent. The ampule was placed in a two-zone furnace and initially heated to 850$^\circ$C for 72 h. A temperature gradient was then established (850$^\circ$C/800$^\circ$C) and maintained for 12 days to facilitate crystal growth, followed by natural cooling to room temperature.

Phase purity and crystal structure were examined by x-ray diffraction (XRD), followed by Rietveld refinement of powder data using FullProf. Crystallographic orientation was confirmed by Laue diffraction. Surface morphology and elemental composition were analyzed by scanning electron microscopy (SEM) and energy-dispersive x-ray spectroscopy (EDX). High-resolution transmission electron microscopy (HRTEM) was used to probe the lattice structure and defects. Electrical transport and penetration-depth measurements were carried out in a cryogen-free magnet system. Magnetization and heat-capacity measurements were performed using a Quantum Design PPMS.

\section{Results and Discussion}

Figure~\ref{fig:xrd}(a) shows the Rietveld refinement of the PXRD pattern of NbSeS. The data are well described by a hexagonal structure with space group $P6_3/mmc$ (No.~194), corresponding to the 2H polytype, similar to the parent compound 2H-NbSe$_2$. The refined lattice parameters are $a = b = 3.393$~\AA\ and $c = 12.336$~\AA, with $\alpha = \beta = 90^\circ$ and $\gamma = 120^\circ$. The chalcogen sites are randomly occupied by Se and S atoms with approximately equal probability. This results in intrinsic substitutional disorder that preserves the global crystal symmetry while introducing local potential fluctuations, which are expected to enhance electronic scattering. Compared with pristine 2H-NbSe$_2$, the reduced $c$-axis lattice parameter reflects the smaller ionic radius of S, leading to a contraction of the interlayer spacing.

Figure~\ref{fig:xrd}(b) shows the XRD pattern of a cleaved single-crystal flake, displaying only sharp $(00l)$ reflections, confirming that the crystallographic $c$ axis is oriented perpendicular to the sample surface~\cite{benz2014introduction}. The absence of additional Bragg peaks indicates high phase purity, while the systematic shift of the $(00l)$ reflections toward higher angles further confirms lattice contraction due to S substitution~\cite{cullity1986elements}. Additionally, the absence of superlattice reflection peaks associated with atomic ordering suggests a random distribution of chalcogen atoms within the lattice~\cite{simonov2020designing,gregory1957elements}. The Laue diffraction pattern [inset (i) of Fig.~\ref{fig:xrd}(b)] exhibits well-defined symmetric spots, indicating high crystalline quality. High-resolution transmission electron microscopy [inset (ii) of Fig.~\ref{fig:xrd}(b)] reveals a well-resolved atomic lattice consistent with the $P6_3/mmc$ symmetry.

Figure~\ref{fig:xrd}(c) presents the EDX spectrum and corresponding elemental maps, which confirm the absence of any impurity phases and a homogeneous distribution of Nb, Se, and S throughout the crystal. The measured atomic contributions (Nb $\approx 33.86\%$, Se $\approx 32.36\%$, and S $\approx 33.78\%$) are consistent with nearly ideal 1:1:1 stoichiometry [table of Fig.~\ref{fig:xrd}(c)]. The uniform spatial distribution of Se and S indicates that disorder is intrinsic and statistically distributed at the atomic scale.

Taken together, these structural characterizations demonstrate that 2H-NbSeS retains its layered hexagonal lattice as its parent compound 2H-NbSe$_2$, highlighting the intrinsic compositional disorder localized within the chalcogen sublattices. This isovalent substitution effectively quenches the long-range CDW instability without altering the nominal valence electron count, thereby preserving the topography of the normal-state Fermi surface. As a result, this system provides an ideal tuning platform to probe the intrinsic superconducting pairing symmetry in the absence of competing coexisting electronic orders.

\begin{figure}
\centering
\includegraphics[width=\columnwidth]{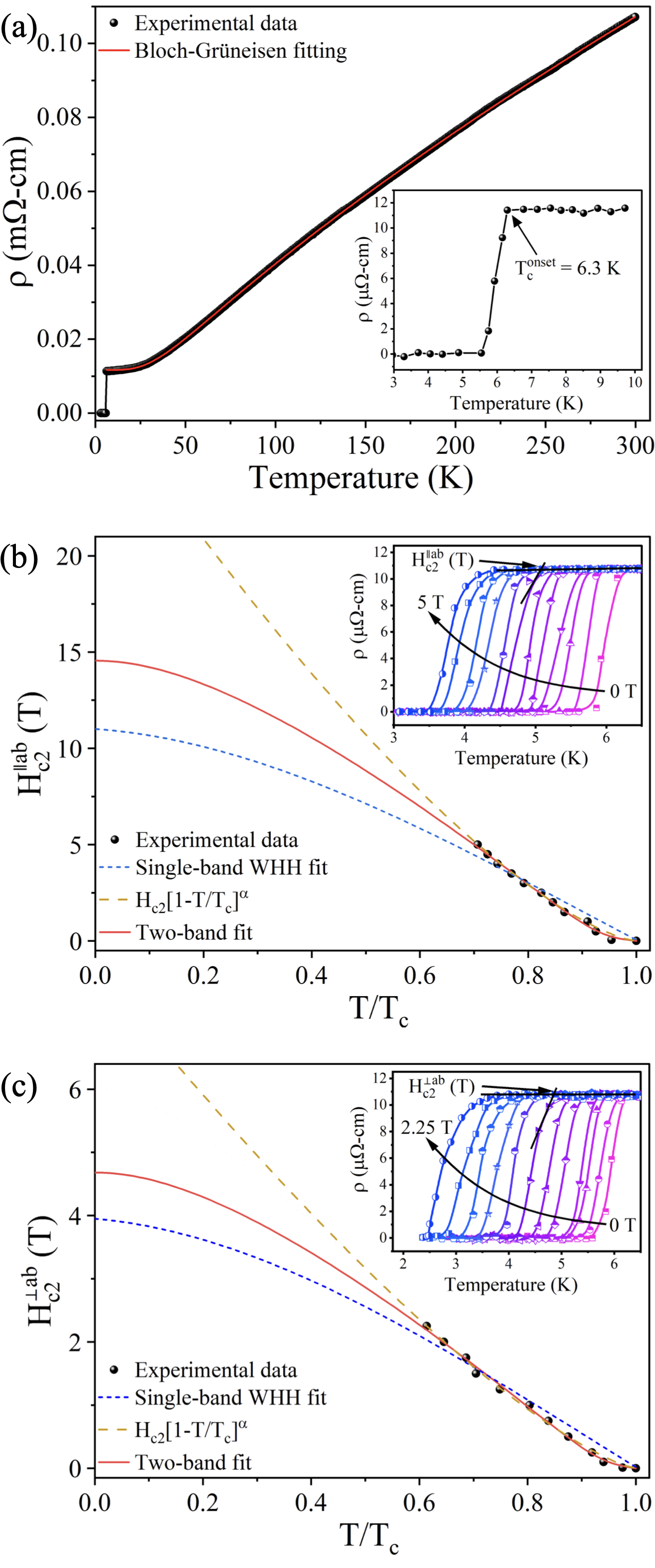}
\caption{(a) Zero-field electrical resistivity as a function of temperature, along with a Bloch-Gr\"uneisen fit in the metallic-state region. The inset shows an enlarged view of the superconducting transition at 6.3 K. (b) and (c) Temperature dependence of the upper critical fields $H_{c2}^{\parallel ab}$ and $H_{c2}^{\perp ab}$ extracted from the field-dependent resistivity data [insets of (b) and (c)]. Solid and dashed curves represent different theoretical fits as described in the text.}
\label{fig:mr}
\end{figure}

Figure~\ref{fig:mr}(a) shows the temperature-dependent electrical resistivity $\rho(T)$ of 2H-NbSeS measured using a standard four-probe configuration. The inset highlights the superconducting transition with an onset temperature $T_c^{\mathrm{onset}} = 6.3$ K. In the normal state, $\rho(T)$ exhibits a nearly linear temperature dependence over a broad temperature range. This indicates the dominance of electron-phonon scattering, with a slight downward curvature as the system approaches $T_c$. Additionally, no signature of CDW order is observed in the resistivity data over the entire measured temperature range, indicating that substitutional disorder suppresses the long-range CDW order present in the parent compound 2H-NbSe$_2$. Such disorder-induced CDW suppression has been reported in several TMD systems, where $T_{\mathrm{CDW}}$ decreases consistently with increasing disorder concentration and eventually vanishes beyond a critical disorder level~\cite{li2017superconducting,cho2018using,mottas2019semimetal}.

To quantify the scattering mechanisms, the normal-state resistivity is analyzed using the Bloch-Gr\"uneisen (BG) model~\cite{gruneisen1933abhangigkeit},
\begin{equation}
\rho(T)=\rho_0+\rho_{BG}(T),
\label{BGmodel}
\end{equation}
where $\rho_0$ represents the residual resistivity arising from static disorder, and $\rho_{BG}(T)$ describes the electron-phonon contribution given by
\begin{equation}
\rho_{BG}(T)=C_{ph}\left(\frac{T}{\Theta_D}\right)^5
\int_0^{\Theta_D/T}\frac{x^5}{(e^x-1)(1-e^{-x})}\,dx.
\label{rhoBG}
\end{equation}
The fit yields $\rho_0 = 11.68~\mu\Omega$-cm, $C_{ph} = 286.41~\mu\Omega$-cm, and a transport Debye temperature $\Theta_D = 195$ K. The residual resistivity ratio $RRR = \rho(300~\mathrm{K})/\rho_0 \approx 9.5$ indicates moderate metallic quality. The relatively large value of $\rho_0$ as compared to pristine 2H-NbSe$_2$~\cite{naito1982galvanomagnetic} reflects enhanced static scattering arising from intrinsic Se/S substitution on the chalcogen sublattice~\cite{kittel2005introduction,mott1936theory}. Such disorder can place the system in a dirty-limit regime, where impurity scattering plays an important role in the superconducting properties~\cite{tinkham1996introduction}. In multiband systems, this can lead to band-selective scattering and unequal electronic diffusivities, which strongly influence the upper critical field and gap structure~\cite{gurevich2003enhancement,hirschfeld2011gap}.

The upper critical fields, $H_{c2}$ were determined from resistivity measurements under applied magnetic fields using the onset criterion. It is defined by the temperature at which the resistivity first deviates from the normal-state behavior [insets of Figs.~\ref{fig:mr}(b) and \ref{fig:mr}(c)]. The resulting temperature dependence of $H_{c2}$ for both field orientations, $H_{c2}^{\parallel ab}$ and $H_{c2}^{\perp ab}$, is shown in Figs.~\ref{fig:mr}(b) and \ref{fig:mr}(c), respectively.

As a first approximation, the data were analyzed using the single-band Werthamer-Helfand-Hohenberg (WHH) model in the dirty limit~\cite{werthamer1966temperature},
\begin{equation}
H_{c2}^{\mathrm{orb}}(0)=-0.693\,T_c\left.\frac{dH_{c2}}{dT}\right|_{T=T_c},
\label{WHH}
\end{equation}
where $\left(dH_{c2}/dT\right)_{T=T_c}$ is the initial slope near $T_c$. While the WHH model reproduces the behavior close to $T_c$, it significantly deviates from the $H_{c2}(T)$ data at lower temperatures and fails to capture the pronounced upward curvature observed for both field orientations. This deviation indicates that a conventional single-band description cannot explain the data and points toward more advanced mechanisms, such as multiband effects and disorder-enhanced scattering. Nevertheless, the WHH expression provides a useful estimate of the orbital limiting field.

Within the weak-coupling BCS framework, the Pauli paramagnetic limiting field is given by~\cite{clogston1962upper,chandrasekhar1962note}
\begin{equation}
H_P \approx 1.84~T_c.
\label{Paulilimit}
\end{equation}
For $T_c=6.3$ K, this yields $H_P \approx 11.6$ T.

To further examine the relative importance of orbital and paramagnetic depairing mechanisms, we estimated the Maki parameter, defined as~\cite{maki1966effect,maki1968critical}
\begin{equation}
\alpha_M=\sqrt{2}\frac{H_{c2}^{\mathrm{orb}}(0)}{H_P},
\end{equation}
where $H_{c2}^{\mathrm{orb}}(0)$ is the orbital limiting field and $H_P$ is the weak-coupling Pauli paramagnetic limit. Using the orbital limiting fields, $H_{c2,~\parallel ab}^{\mathrm{orb}}(0)\approx 11.06$~T and $H_{c2,~\perp ab}^{\mathrm{orb}}(0)\approx 3.94$~T, estimated from the WHH expression, we obtain
\begin{equation}
\alpha_M^{\parallel ab}\approx 1.34,
\qquad
\alpha_M^{\perp ab}\approx 0.48 .
\end{equation}
The different values of $\alpha_M$ for the two field configurations reflect the anisotropic nature of pair-breaking mechanisms in 2H-NbSeS. For $H\parallel ab$, the relatively large Maki parameter indicates that Pauli paramagnetic pair breaking becomes important for the in-plane field configuration. In contrast, the smaller value obtained for $H\perp ab$ indicates that superconductivity in the out-of-plane field configuration remains predominantly governed by orbital depairing.

The large value of $\alpha_M^{\parallel ab}$, together with the upward curvature of $H_{c2}(T)$, suggests that the upper critical field behavior in 2H-NbSeS is governed not by a single mechanism, but by a combined effect of intrinsic disorder, multiband superconductivity, and spin-paramagnetic limitation~\cite{lee2009effects,xing2017two,zocco2013pauli}. This interpretation is consistent with the large diffusivity imbalance obtained from the dirty two-band analysis. The intrinsic Se/S substitutional disorder introduces additional scattering that can influence different electronic bands unequally, thereby modifying the orbital limiting field and superconducting anisotropy in the multiband state~\cite{askerzade2012unconventional}.

To describe the deviations from conventional GL behavior near $T_c$, we analyzed the data using a phenomenological scaling form~\cite{freudenberger1998superconductivity},
\begin{equation}
H_{c2}(T)=H_{c2}(0)
\left[1-\left(\frac{T}{T_c}\right)\right]^{\alpha},
\label{Hc2phenomenological}
\end{equation}
which yields exponents $\alpha=1.47$ for $H\parallel ab$ and $\alpha=1.32$ for $H\perp ab$. The values $\alpha>1$ indicate a stronger suppression of superconductivity with applied field than predicted by conventional Ginzburg-Landau theory. This behavior is commonly observed in systems with anisotropy, disorder-induced scattering, and multiband superconductivity~\cite{kneidinger2013superconductivity,liu2021enhanced,yang2023discovery}.

The pronounced upward curvature of $H_{c2}(T)$ reflects interband coupling and band-dependent scattering in the dirty limit~\cite{shulga1998upper}. To describe this behavior, the data were analyzed within a two-band model formulated in the diffusive (Usadel) framework~\cite{gurevich2003enhancement},
\begin{align}
a_0[\ln t+U(h)][\ln t+U(\eta h)] + a_1[\ln t+U(h)] \notag \\
+ a_2[\ln t+U(\eta h)] = 0,
\label{Hc2twoband}
\end{align}
where $t=T/T_c$, $h=HD_1/(2\varphi_0T)$, $\eta=D_2/D_1$, and $U(x)=\psi(x+1/2)-\psi(1/2)$. The coefficients depend on the intraband and interband coupling constants $\lambda_{ij}$.

The two-band model provides an excellent fit to the experimental data for both field orientations. For $H\parallel ab$, the fitting parameters are $\lambda_{11}=1.53$, $\lambda_{22}=0.87$, and $\lambda_{12}=\lambda_{21}=0.29$, with a diffusivity ratio $\eta=6.17$. For $H\perp ab$, the corresponding parameters are $\lambda_{11}=1.22$, $\lambda_{22}=1.09$, and $\lambda_{12}=\lambda_{21}=0.14$, with $\eta=7.85$.

A key result of this analysis is the large diffusivity ratio $\eta\gg 1$, which provides strong evidence for band-selective scattering. This implies that scattering affects different Fermi-surface sheets unequally~\cite{kim2012measurement,gozzelino2009intraband}. These differences contribute to both the enhancement and curvature of $H_{c2}(T)$. Furthermore, the intraband coupling constants $\lambda_{11}$ and $\lambda_{22}$ are larger than the interband couplings $\lambda_{12}$ and $\lambda_{21}$. This indicates that superconducting pairing mainly develops within individual bands. The finite interband coupling ensures a single superconducting transition temperature, while the different diffusivities and coupling strengths give rise to multiple superconducting energy bands~\cite{suhl1959bcs,moskalenko1959superconductivity,gurevich2003enhancement}. These results suggest that intrinsic substitutional disorder in 2H-NbSeS promotes band-selective multiband superconducting behavior in the dirty limit.

The extrapolated zero-temperature upper critical fields are
\begin{equation}
H_{c2}^{\parallel ab}(0)=14.55~\mathrm{T}, \qquad
H_{c2}^{\perp ab}(0)=4.68~\mathrm{T}.
\label{Hc2twobandvalues}
\end{equation}
Within anisotropic Ginzburg-Landau (GL) theory, the upper critical fields are related to the superconducting coherence lengths $\xi$ by~\cite{tinkham1996introduction,klemm2012layered}
\begin{equation}
H_{c2}^{\perp ab}=\frac{\Phi_0}{2\pi \xi_{ab}^2},\qquad
H_{c2}^{\parallel ab}=\frac{\Phi_0}{2\pi \xi_{ab}\xi_c},
\label{coherencelength}
\end{equation}
where $\Phi_0=2.07\times10^{-15}$ Wb is the magnetic flux quantum, $\xi_{ab}$ and $\xi_c$ are the coherence lengths in $ab$-plane and $c$-axis, respectively.

Using $H_{c2}^{\perp ab}(0)$ and $H_{c2}^{\parallel ab}(0)$ the coherence lengths are obtained as
\begin{equation}
\xi_{ab}\approx 8.4~\mathrm{nm},\qquad
\xi_c\approx 2.7~\mathrm{nm}.
\label{coherencelengthvalues}
\end{equation}
The resulting anisotropy parameter is
\begin{equation}
\gamma=\frac{\xi_{ab}}{\xi_c}\approx 3.1.
\label{anisotropy}
\end{equation}
The moderate anisotropy reflects the quasi-two-dimensional electronic structure of the 2H layered system, arising from comparatively weaker interlayer coupling along the $c$-axis, as commonly observed in layered superconductors~\cite{prober1980upper}.

\begin{figure}
\centering
\includegraphics[width=\columnwidth]{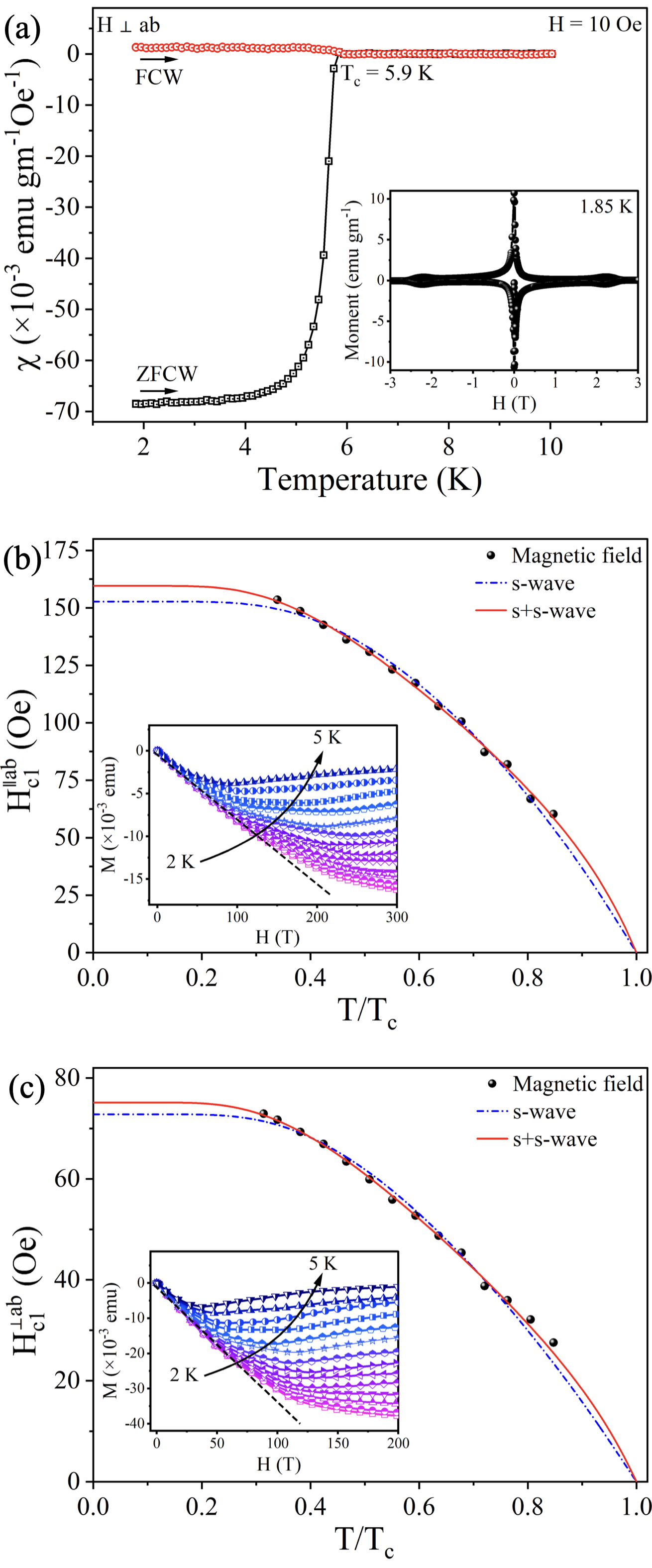}
\caption{(a) Temperature-dependent magnetization (ZFCW-FCW) measured in an applied field of 10 Oe with $H\perp ab$. The inset shows the field-dependent magnetization ($M$-$H$ loop) at 1.85 K. (b) and (c) Temperature dependence of the lower critical field $H_{c1}(T)$ extracted for $H\parallel ab$ and $H\perp ab$, respectively [insets of (b) and (c)]. The dash-dotted blue lines correspond to single-gap $s$-wave fits and the solid red lines to two-gap $s+s$ wave fits.}
\label{fig:mh}
\end{figure}

The superconducting transition in 2H-NbSeS is further confirmed by dc magnetization measurements. Fig.~\ref{fig:mh}(a) shows the temperature dependence of the magnetic susceptibility measured with an applied field of 10 Oe with $H\perp ab$ for both zero-field-cooled warming (ZFCW) and field-cooled warming (FCW) conditions. A sharp diamagnetic transition is observed at $T_c\approx 5.9$ K. This value is slightly lower than the transition temperature obtained from resistivity, reflecting the difference between bulk diamagnetic response and the onset of percolative superconducting paths~\cite{harkonen1989percolation}. The large shielding fraction confirms bulk superconductivity, and the separation between the ZFCW and FCW curves indicates significant vortex pinning, which is an essential characteristic of a type-II superconductor~\cite{blatter1994vortices}. The inset of Fig.~\ref{fig:mh}(a) shows the isothermal $M$-$H$ loop measured at 1.85~K. A slight jump in magnetization appears around $H \approx 2.2$~T. Such changes are commonly referred to as flux jumps, which originate from sudden vortex redistribution caused by strong pinning~\cite{chikurov2021magnetic}.

The temperature-dependent lower critical field $H_{c1}(T)$ for both field orientations is shown in Figs.~\ref{fig:mh}(b) and \ref{fig:mh}(c). The lower critical field $H_{c1}(T)$ is determined from the field at which $M$ deviates from its linear Meissner response [insets of Figs.~\ref{fig:mh}(b) and \ref{fig:mh}(c)]. As $H_{c1}(T)$ is directly proportional to the superfluid density $\rho_s(T)$, it provides a sensitive probe to determine the superconducting gap structure. The lower critical field at temperature $T$ can be expressed as~\cite{zhou2025nodeless}
\begin{equation}
H_{c1}(T) = H_{c1}(0)\,\rho_s(T),
\label{Hc1swave}
\end{equation}
where $H_{c1}(0)$ denotes the lower critical field at zero temperature.

For a single isotropic $s$-wave gap, the normalized superfluid density is given by~\cite{zhou2025nodeless}
\begin{equation}
\rho_s(T) = 1 + 2 \int_{\Delta(T)}^{\infty} 
\frac{\partial f}{\partial E}
\frac{E\, dE}{\sqrt{E^2 - \Delta^2(T)}},
\label{Hc1superfluid}
\end{equation}
where $f(E,T)=\left[1+\exp\left(E/k_B T\right)\right]^{-1}$ is the Fermi-Dirac distribution function. The temperature dependence of the superconducting gap is approximated by~\cite{zhou2025nodeless}
\begin{equation}
\Delta(T) = \Delta_0 
\tanh \left[ 1.82 \left(1.018\left(\frac{1}{t}-1\right)\right)^{0.51} \right],
\label{deltaT}
\end{equation}
with $t = T/T_c$, which accurately reproduces the BCS gap evolution over the full temperature range. The blue dash-dotted line represents the single-band $s$-wave fit, which fails to reproduce the experimental data.

To examine possible multiband effects, the data were also fitted using a two-band $s+s$ wave model, in which the total superfluid density is written as
\begin{equation}
\rho_s(T) = x_1\rho_1(T) + x_2\rho_2(T),
\label{splussHc1}
\end{equation}
where $\rho_1(T)$ and $\rho_2(T)$ correspond to the superfluid densities associated with the larger ($\Delta_1$) and smaller ($\Delta_2$) band gaps, respectively, and $x_1$ and $x_2$ denote the fractional contributions of the two bands. The red solid line represents the two-band $s+s$ wave fit to the experimental data.

For $H \parallel ab$, the two-band fit yields
\begin{equation}
2\Delta_1/k_B T_c = 3.98, \qquad
2\Delta_2/k_B T_c = 1.60,
\label{bandvalueHpara}
\end{equation}
with weight factors $x_1 = 0.80$ and $x_2 = 0.20$.
For $H \perp ab$, the extracted parameters are
\begin{equation}
2\Delta_1/k_B T_c = 4.00, \qquad
2\Delta_2/k_B T_c = 1.58,
\label{bandvalueHper}
\end{equation}
with $x_1 = 0.84$ and $x_2 = 0.16$. The nearly identical gap magnitudes and weights obtained for both field orientations indicate that the multiband superconductivity is intrinsic rather than field-direction-dependent.

Compared with the single-band $s$-wave model, the two-band $s+s$ wave model reproduces the experimental $H_{c1}(T)$ data over the entire temperature range with better fit, particularly at low temperatures. This analysis supports the presence of two distinct superconducting gaps and provides strong evidence for multiband superconductivity in 2H-NbSeS. The larger gap, $2\Delta_1/k_B T_c \approx 4.0$, slightly exceeds the weak-coupling BCS value of 3.53~\cite{tinkham1996introduction}, suggesting moderately enhanced coupling on one Fermi-surface sheet, whereas the smaller gap is significantly suppressed. Such gap asymmetry is characteristic of multiband superconductivity in the presence of band-dependent scattering~\cite{romeo2015minimal}.

From the analysis, the zero-temperature lower critical fields are obtained as $H_{c1}^{\parallel ab}(0)=169.65$ Oe and $H_{c1}^{\perp ab}(0)=75.10$ Oe. Within Ginzburg-Landau theory, the lower critical field for $H\perp ab$ is related to the penetration depth $\lambda_{ab}(0)$ and coherence length $\xi_{ab}(0)$ by
\begin{equation}
H_{c1}^{\perp ab}(0)=\frac{\Phi_0}{4\pi \lambda_{ab}^2(0)}\ln\kappa_c,
\label{penetration_depth}
\end{equation}
where $\kappa_c=\lambda_{ab}(0)/\xi_{ab}(0)$ is the GL parameter. Using $\xi_{ab}(0)\approx 8.40$ nm, the values of $\lambda_{ab}(0)$ and $\kappa_c$ are obtained as $\lambda_{ab}(0)\approx 276.9$ nm and $\kappa_c\approx 32.9$. Using the anisotropy parameter $\gamma\approx 3.11$, the out-of-plane penetration depth is estimated as $\lambda_c(0)\approx 861.1$ nm.

\begin{figure}
\centering
\includegraphics[width=\columnwidth]{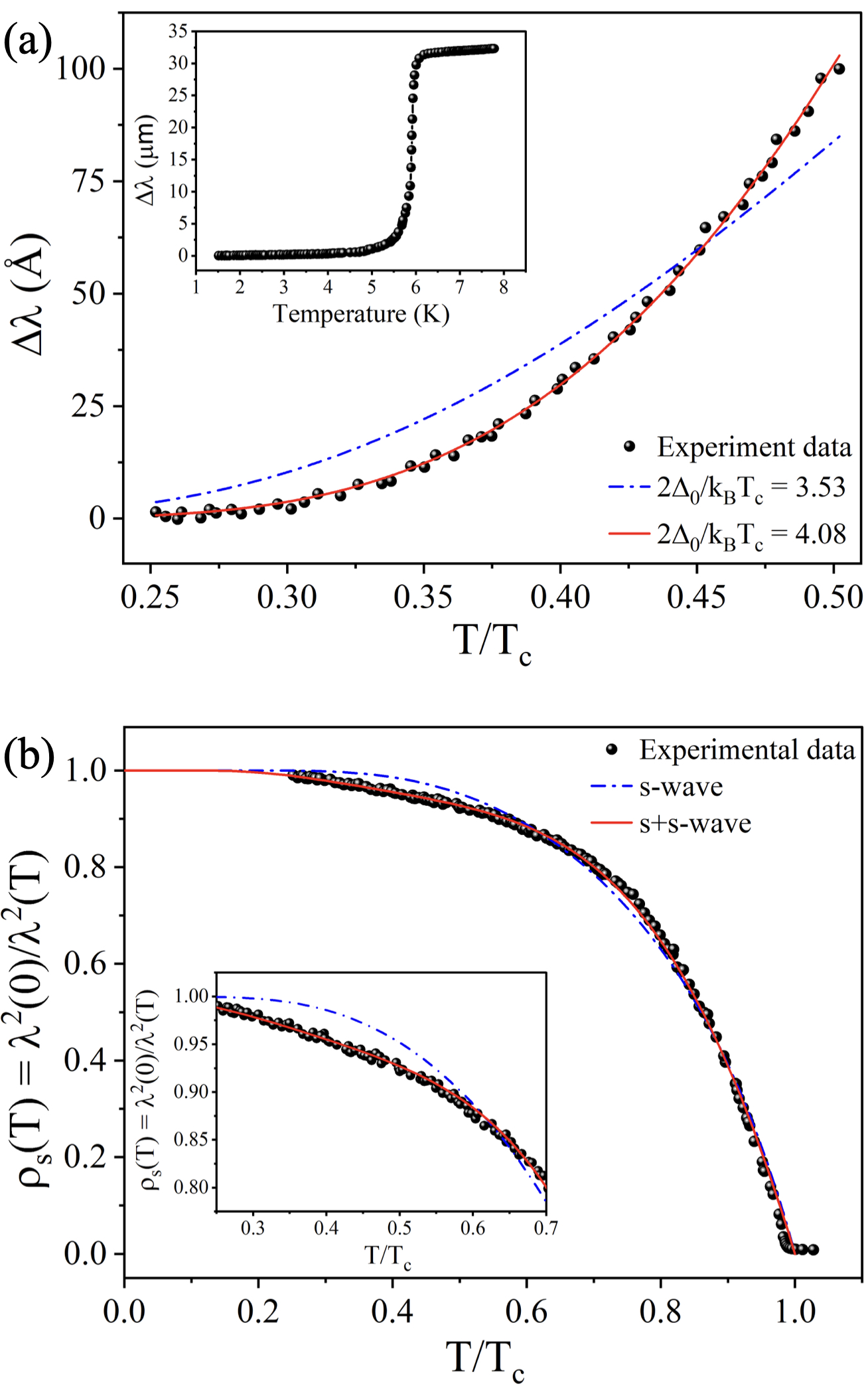}
\caption{(a) Low-temperature penetration-depth variation $\Delta\lambda(T)$ as a function of reduced temperature down to $0.25\,T_c$, fitted with exponential models [solid red line with $2\Delta_0/k_B T_c=4.08$ and dot-dashed blue line with $2\Delta_0/k_B T_c=3.53$]. The inset shows the variation of the rf penetration depth as a function of temperature in zero applied magnetic field. (b) Superfluid density $\rho_s(T)=\lambda^2(0)/\lambda^2(T)$ derived from $\lambda(T)$, fitted over the full temperature range using both single-gap and $s+s$ wave models. The inset highlights the low-temperature region.}
\label{fig:tdo}
\end{figure}

To gain further insight into the superconducting pairing symmetry, the temperature dependence of the magnetic penetration depth $\Delta\lambda(T)$ was measured using the tunnel diode oscillator (TDO) technique~\cite{prozorov2006magnetic}. This technique is highly sensitive to small changes in the superconducting screening response. The measured frequency shift $\Delta F$ was converted into the change in penetration depth $\Delta\lambda(T)$ using a system-dependent calibration factor $G = 2.27$\AA/Hz, such that $\Delta\lambda(T) = G\Delta F$~\cite{yadav2025possible}. The inset of Fig.~\ref{fig:tdo}(a) shows $\Delta\lambda(T)$ in zero applied magnetic field, exhibiting a superconducting transition at $T_c\approx 6.0$ K.

At low temperatures, the penetration depth of a fully gapped superconductor follows an exponential temperature dependence~\cite{kim2013magnetic},
\begin{equation}
\Delta\lambda(T)=\lambda(0)
\sqrt{\frac{\pi\Delta_0}{2k_B T}}
\exp\!\left(-\frac{\Delta_0}{k_B T}\right),
\label{fig:tdopentration}
\end{equation}
which is characteristic of isotropic $s$-wave pairing. A fit using the weak-coupling BCS value $2\Delta_0/k_B T_c=3.53$ fails to reproduce the experimental data. However, an enhanced gap ratio $2\Delta_0/k_B T_c=4.08$ provides an excellent fit of the low-temperature behavior. This enhancement indicates that the dominant superconducting gap exceeds the weak-coupling limit, consistent with moderately strengthened pairing on one Fermi-surface sheet.

It is important to note that in a multiband superconductor, the low-temperature behavior of $\Delta\lambda(T)$ can often be described by an effective gap scale associated with the dominant superconducting band. The extracted value therefore represents an effective gap rather than the full multiband gap structure~\cite{carrington2003penetration,hashimoto2009microwave}. 

To probe the full superconducting gap structure, the normalized superfluid density $\rho_s(T) = [\lambda(0)/\lambda(T)]^2$ was determined using experimental data for the temperature-dependent penetration depth~\cite{kim2013magnetic}. The superfluid data was analysed using Eqs.~(\ref{Hc1superfluid}) and~(\ref{deltaT}). While a single-gap $s$-wave model reproduces the behavior near $T_c$, it fails to account for the systematic deviation at low temperatures. In contrast, a two-band $s+s$ wave model, Eq.~(\ref{splussHc1}), provides an excellent description of the data, as shown in Fig.~\ref{fig:tdo}(b).

The best fit yields two isotropic superconducting gaps,
\begin{equation}
2\Delta_1(0)/k_B T_c=4.24,\qquad
2\Delta_2(0)/k_B T_c=1.46,
\label{bandvalueTDO}
\end{equation}
with relative weights $x_1=0.85$ and $x_2=0.15$. The large disparity between the two gap magnitudes indicates the presence of two different bands in the superconducting state. The consistency between the penetration-depth analysis and the large diffusivity imbalance obtained from the $H_{c2}$ measurements further supports multiband superconductivity in 2H-NbSeS.

\begin{figure}
\centering
\includegraphics[width=\columnwidth]{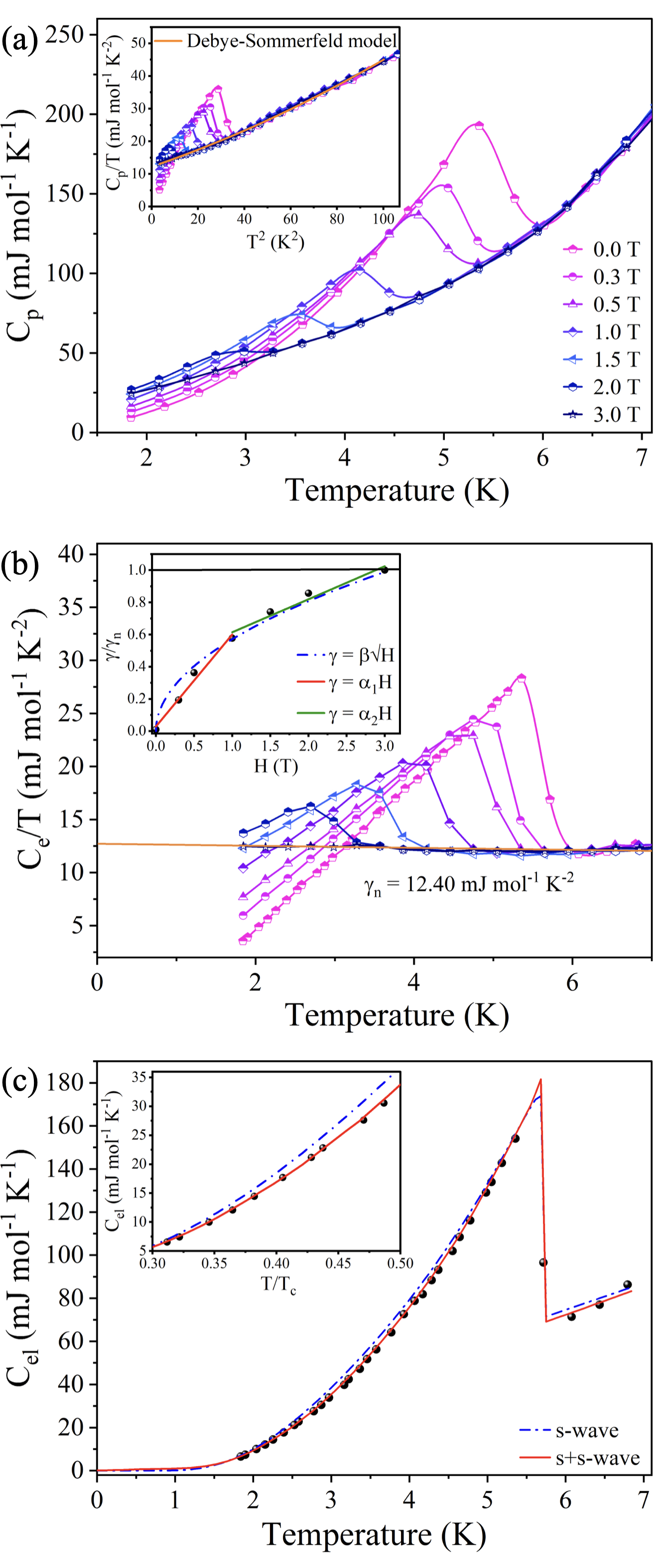}
\caption{(a) Temperature dependence of the specific heat $C_p(T)$ measured under various magnetic fields. The inset shows $C_p/T$ versus $T^2$ along with the Debye-Sommerfeld fit (solid orange line). (b) $C_e/T$ versus $T$ at different magnetic fields; the inset shows the field dependence of the Sommerfeld coefficient $\gamma(H)$. (c) Temperature dependence of the electronic specific heat $C_e(T)$ at zero field, fitted with single-gap $s$-wave and two-gap $s+s$ wave models.}
\label{fig:hc}
\end{figure}

To probe the bulk thermodynamic response of the electronic state, heat capacity measurements were carried out on single crystal flake of 2H-NbSeS. These measurements provide a direct probe of the electronic density of states and allow quantitative determination of superconducting parameters such as the energy gap and coupling strength~\cite{prakash2015multiband}. They also offer an important thermodynamic signature of the superconducting transition, independent of transport or magnetic measurements. Fig.~\ref{fig:hc}(a) shows the temperature-dependent specific heat of 2H-NbSeS measured under various applied magnetic fields. The data exhibit a clear anomaly at the superconducting transition, confirming the bulk nature of superconductivity.

The inset of Fig.~\ref{fig:hc}(a) shows that the normal-state specific heat was analyzed using the Debye-Sommerfeld model~\cite{balakrishnan2013superconducting},
\begin{equation}
C_p(T)=\gamma_n T+\beta T^3+\delta T^5,
\label{Debye-Sommerfeld model}
\end{equation}
yielding $\gamma_n=12.40~\mathrm{mJ\,mol^{-1}K^{-2}}$, $\beta=4.42\times10^{-1}~\mathrm{mJ\,mol^{-1}K^{-4}}$, and $\delta=8.46\times10^{-3}~\mathrm{mJ\,mol^{-1}K^{-6}}$. The Debye temperature extracted from $\beta$ is $\Theta_D\approx 236$ K, which is consistent with the value obtained from transport measurements.

After subtracting the lattice contribution, the electronic specific heat $C_e/T$ is presented in Fig.~\ref{fig:hc}(b). The electronic specific heat was analyzed using a phenomenological exponential form~\cite{zhou2025nodeless},
\begin{equation}
\frac{C_e}{T} = \gamma + \frac{a}{T} \exp\left(-\frac{b T_c}{T}\right),
\label{eq:sommerfeld}
\end{equation}
where $\gamma$ is the residual Sommerfeld coefficient, and $a$ and $b$ are fitting parameters. The low-temperature data at each magnetic field were fitted using Eq.~(\ref{eq:sommerfeld}), and the fits were extrapolated to $T = 0$~K to extract the Sommerfeld coefficient $\gamma$. The inset of Fig.~\ref{fig:hc}(b) shows the resulting field dependence of $\gamma(H)$.

The observed behavior deviates from both the linear field dependence expected for a single-band fully gapped superconductor and the $\sqrt{H}$ dependence characteristic of nodal superconductivity. Instead, $\gamma(H)$ exhibits two distinct regimes, with a rapid increase at low fields followed by a slower variation at higher fields. This nontrivial field dependence reflects the presence of multiple superconducting bands with different energy scales and field sensitivities, a hallmark of multiband superconductivity~\cite{zhou2025nodeless}.

The temperature dependence of the electronic specific heat $C_e(T)$ at zero field is shown in Fig.~\ref{fig:hc}(c), with a superconducting transition temperature $T_c=5.7$ K obtained from entropy conservation. The temperature dependence of the electronic specific heat in the superconducting state can be described within a fully gapped model. The normalized entropy is given by~\cite{bouquet2001phenomenological}
\begin{equation}
\frac{S}{\gamma_n T_c} = -\frac{6}{\pi^2} \frac{\Delta(0)}{k_B T_c}
\int_0^{\infty} \left[ f \ln f + (1-f)\ln(1-f) \right] dy,
\label{entropy}
\end{equation}
where $f(E,T) = \left[\exp\left(E/k_B T\right) + 1\right]^{-1}$ is the Fermi-Dirac distribution function, $E = \sqrt{\xi^2 + \Delta^2(T)}$ is the quasiparticle energy, and $\xi$ is the normal-state electron energy relative to the Fermi level. Here, $y = \xi/\Delta(0)$ and $t = T/T_c$. The normalized electronic specific heat is then obtained from the entropy as
\begin{equation}
\frac{C_e}{\gamma_n T_c} = t \frac{d(S/\gamma_n T_c)}{dt}.
\label{specificheat}
\end{equation}
To probe the superconducting gap structure, the electronic specific heat was analyzed within both single-gap and two-band models. While a single-gap $s$-wave description reproduces the behavior near $T_c$, it fails to account for the excess low-temperature quasiparticle contribution. In contrast, a two-band $s+s$ wave model provides an excellent fit over the entire temperature range.

The best fit yields two distinct superconducting gaps,
\begin{equation}
2\Delta_1(0)/k_B T_c=3.68,\qquad
2\Delta_2(0)/k_B T_c=1.28,
\label{bandvalueheatcapacity}
\end{equation}
with relative weights $x_1=0.87$ and $x_2=0.13$. The large disparity between the two gap magnitudes indicates a strongly asymmetric multiband superconducting state.

\begin{table}[t]
\caption{\label{tab:gaps}Comparison of superconducting gap parameters obtained from different experimental probes for 2H-NbSeS.}
\begin{ruledtabular}
\begin{tabular}{lcccc}
Probe & $2\Delta_1/k_B T_c$ & $2\Delta_2/k_B T_c$ & Weight 1 & Weight 2 \\
\hline
$H_{c1}(T)$, $H\parallel ab$ & 3.98 & 1.60 & 0.80 & 0.20 \\
$H_{c1}(T)$, $H\perp ab$     & 4.00 & 1.58 & 0.84 & 0.16 \\
$\rho_s(T)$ from TDO         & 4.24 & 1.46 & 0.85 & 0.15 \\
$C_e(T)$                     & 3.68 & 1.28 & 0.87 & 0.13 \\
\end{tabular}
\end{ruledtabular}
\end{table}

The gap values obtained from specific heat are in good agreement with those extracted from lower critical field and penetration-depth measurements (Table~\ref{tab:gaps}), demonstrating the robustness of the two-gap scenario across multiple experimental probes. The large diffusivity imbalance obtained from $H_{c2}$ measurements, together with the strong gap asymmetry observed in $C_e(T)$, $H_{c1}(T)$, and $\lambda(T)$, provides evidence for band-selective scattering. 

The present study provides important insight into the nature of superconducting pairing in the NbSe$_2$ family. In 
2H-NbSe$_2$, coexisting CDW order can mimic a two-gap structure in tunneling and thermodynamic measurements, making it difficult to distinguish genuine multiband pairing from a CDW-driven artifact. The absence of CDW order in 2H-NbSeS removes this ambiguity entirely. The two distinct superconducting gaps observed here cannot be attributed to any CDW-related effect and must therefore arise from intrinsic multiband pairing on the multi-sheet Nb-$4d$ Fermi surface. Since the superconducting condensate in both 2H-NbSe$_2$ and 2H-NbSeS is predominantly governed by the Nb-$4d$-derived Fermi-surface sheets rather than the chalcogen $p$-derived pocket~\cite{yokoya2001fermi}, the robust two-gap behavior observed here in the absence of CDW order suggests that multiband superconductivity in the NbSe$_2$ family is intrinsic rather than any CDW derived artifact.

\section{Conclusion}

We have investigated the superconducting properties of single-crystalline 2H-NbSeS, a mixed-chalcogen analogue of 2H-NbSe$_2$ in which random Se/S substitution introduces intrinsic disorder and suppresses the CDW order present in the parent compound. Transport, magnetic, and thermodynamic measurements consistently reveal two distinct nodeless superconducting gaps with moderate anisotropy ($\gamma \approx 3$). The large diffusivity imbalance obtained from the dirty-limit two-band analysis of $H_{c2}(T)$, combined with the two-gap signatures observed in $H_{c1}(T)$, superfluid density, and electronic specific heat, consistently establish 2H-NbSeS as a multiband superconductor. Importantly, because long-range CDW order is absent in 2H-NbSeS, the observed two-gap behavior cannot be attributed to CDW-induced gap modulations. Instead it strongly supports an intrinsic multiband origin of superconductivity. These results support the view that multiband superconductivity in the NbSe$_2$ family originates from its underlying electronic structure and not due to CDW correlations.

\begin{acknowledgments}
K. Y. acknowledges the Council for Scientific and Industrial Research (CSIR) for a Senior Research Fellowship. The authors gratefully acknowledge the Department of Science and Technology, Government of India, for the low-temperature and high-field facility at JNU (FIST program). We also thank Anusandhan National Research Foundation
(Grant No. ANRF/PAIR/2025/000029/PAIR-A) and the Nano-Mission project (DST/NM/TUE/QM-10/2019(G)/6) for chemicals, consumables and equipment grants. 
\end{acknowledgments}

\bibliography{NbSeS}

\end{document}